\documentclass[aps,prb,twocolumn,preprintnumbers,amsmath,amssymb]{revtex4}
\usepackage{dcolumn}
\usepackage{bm}
\usepackage{amsmath}
\usepackage{feynmp}
\usepackage{graphicx}

\begin{document}
\title{Electronic Conduction in Disordered Materials: Wave-like or Particle-like?}
\author{Vincent E. Sacksteder IV$^{1}$}
\affiliation{W155 Wilson Building, Royal Holloway University of London, Egham Hill, Egham, TW20 0EX, United Kingdom }
\email{vincent@sacksteder.com}

\date{\today}

\begin{abstract}
When materials are in the  diffusive regime, i.e. when the scattering length is less than the sample size, charge transport is mediated by diffusons and cooperons.  We argue that these charge carriers are not always spread out throughout the sample, like waves.  When the scattering length is smaller than the sample size, diffusons and cooperons follow trajectories inside the sample which have an effective width determined by the scattering length.    The net effect is particle-like motion.  We discuss three experiments whose results confirm the existence of cooperons  following particle-like trajectories.  One experiment, with magnetic field perpendicular to a topological insulator surface, resolved and measured the areas of  Cooperon loops.  The other two experiments, with magnetic field parallel to wires where conduction occurred on the wire surface, measured the number of times that a Cooperon trajectory wound around the wire.
\end{abstract}

\maketitle

In clean [ballistic] materials  electronic states are described by Bloch waves, functions with the same infinite lattice symmetry as the underlying material, extending infinitely throughout the material.  The physical picture of the motion is unimpeded electronic motion along a straight line without any limit.   However the electron is clearly a wave - its wave-function exists everywhere.

In disordered materials scattering centers break the lattice symmetry.   Perturbative formalisms describe the resulting motion as sequences of straight line motion, followed by a scattering event, followed by straight line motion, followed by a scattering event, and so on. \cite{RevModPhys.69.731,RevModPhys.58.323,RevModPhys.57.287,mirlin2000statistics}
  At each scattering the electron's direction, phase, and possibly also energy [if the disorder is time-dependent] are altered.  The trajectories described by the wave-function therefore look like sequences of straight-line segments.  Nonetheless the wave-function continues to be built from Bloch waves, i.e. it is a superposition of Bloch waves.  This implies that the electron is wave-like, extended throughout the entire material.

This picture breaks down in samples whose length $L$ exceeds the localization length $L_{LOC} \approx l N$, where $N$ is the number of conducting channels and $l$ is the scattering length.  At this scale Anderson localization causes the electronic wave-function to remain localized within a region of size $L_{LOC}$, and outside of this region the electron does not penetrate.  In this case the Bloch-wave picture of extended wave-like electrons completely fails.

This paper is concerned neither with clean [ballistic] materials nor with Anderson-localized materials, but instead with the intermediate regime, called the diffusive regime. In this regime scattering randomizes the phase of individual electrons and holes, impeding their transmission of charge.  It is only when an electron and hole traverse the same sequence of scatterings that they are able to transmit charge.   If the electron and hole follow the scatterings in the same order, they compose into the diffuson which mediates classical conduction.  If the electron and hole follow the same scatterings but in orders that are reversed with respect to each other, then they compose into the cooperon, which manifests quantum interference's ability to increase or decrease probabilities above or below their classically expected values.  The cooperon mediates quantum interference effects such as weak localization, weak antilocalization (WL/WAL), and universal conductance fluctuations (UCFs).      In other words, diffusons and cooperons, both of which are electron-hole pairs induced by scattering, are the key to conduction in diffusive systems.

The main point of this paper is that, in a specific sample with scattering centers located at specific sites, it may be inappropriate  to think that individual diffusons and cooperons are extended throughout the sample.  Instead, each diffuson and cooperon may follow a specific trajectory determined by the scattering centers.  These trajectories are tube-like, with width determined by the scattering length, and they change direction at each scattering center.  In this respect the diffuson and cooperon  are more like particles than like waves.

In particular, we discuss one particular experiment on the conduction in a TI nanowire, whose results can have no other explanation than that charge conduction on the surface of that TI wire was localized on a path which wound around the wire.  The path had a well-defined winding number, i.e. it wound around the wire a specific integer number of times.  The presence of this winding number in the experimental data is clear proof that the electrons did not occupy the entire surface of the wire, but instead followed a trajectory around the wire.

I am not questioning the validity of the standard  formalism \cite{RevModPhys.69.731,RevModPhys.58.323,RevModPhys.57.287,mirlin2000statistics}
 for calculating disordered conduction, and I do not  disagree with its results, which include weak localization and UCFs.   I am rather questioning the interpretation of these results.   The standard formalism starts its work by averaging over the ensemble of all possible samples, i.e. all possible alternative locations of the scattering centers.  Therefore its results always pertain to the average position of an electron, as if one had an infinite number of samples, measured all of them, and then found the average position of the electrons, their average conductance, etc.  Sample-to-sample variation can be probed by calculating higher moments of observables, but this method is rarely taken beyond calculating the second moment, i.e. the standard deviation, and rare events known as anomalously localized states.   The standard approach based on ensemble averaging cannot avoid predicting that the electrons extend throughout the entire sample.

In contrast, our focus here is on individual samples.  Each sample has a unique configuration of scattering centers, and these guide the motion of diffusons and cooperons.  The mechanism which allows these electron-hole pairs to conduct over long distances is phase cancellation - the random phase incurred when the electron meets a scatterer is cancelled by an opposite phase incurred by the hole when it meets the same scatterer.    The phase incurred by  motion between scatterers also cancels, as long as both the electron and the hole follow the same line segments connecting scatterers into a trajectory through the sample.  However when the electron and hole stray from too far from each other, they no longer guard each other's phase.  

The relevant length scale over which electron and hole protect each other, which sets the effective width of the diffuson or cooperon trajectory, is the decay length of the [ensemble-averaged] single-electron wavefunction.  This can easily be verified by calculating the bubble diagram that mathematically describes diffuson and cooperons and analyzing the length scales involved.   The long length scale that allows  diffusons and cooperons to conduct over long distances is determined by a special cancellation of terms when the electron and hole are moving in synchrony.  In contrast, since motion of the electron and hole towards or away from each other is not accompanied by any cancellation of terms, it simply the inherits the short length scale of the ensemble-averaged Green's functions that contribute to the cooperon/diffuson bubble diagram.  In other words the width of the diffuson or cooperon is set by the decay length of the ensemble-averaged Green's function, which is  tied to the decay length of the ensemble-averaged wavefunctions, and also to the material's scattering length $l$.  Therefore diffusons and cooperons follow paths which  have an effective width proportional to $l$.  As long as $l$ is smaller than the sample size, the diffusons and cooperons mediating charge conduction act more like particles than like waves.

\subsection{Universal Conductance Fluctuations}
We distinguish between several factors that control which paths are taken  by diffusons and cooperons:
\begin{itemize}
\item Placement and type of scattering centers.  Trajectories are deterministic over short distances but become chaotic over longer distances, as more and more scatterings occur, similar to motion in a pinball machine or in an chaotic billiard.  This causes sample-unique fingerprints visible in measurements of electronic transport, called universal conductance fluctuations (UCFs).    
\item The Fermi level.  Electronic conduction is determined by  the electrons and holes near the Fermi level.  Their energy determines their speed, and therefore determines the precise outcome of each scattering event, which in turn determines the path that is taken.   The paths taken by diffusons and cooperons depend sensitively on the Fermi level.
\item The magnetic field, via the Zeeman effect.  This effect splits and changes the electron and hole energies and therefore influences electron trajectories in a way that is in some respects analogous to the effect of the Fermi level. In materials where the spin relaxation length is comparable to the scattering length, such as topological insulators, this effect has a negligible influence on diffusive electronic transport. \cite{PhysRevB.83.241304} In such materials charge is transmitted but not spin polarization, and both diffusons and cooperons are reduced to their spin singlet components.
\end{itemize}

The transport experiments discussed in this article incorporate the above factors, but focus on one more very powerful experimental tool for understanding disordered transport: the phase factor induced by putting a sample in a magnetic field.   If magnetic field flows through a sample it adds a magnetic phase $\exp(\imath \Phi)$ to cooperon [but not diffuson] paths, when they trace loops returning to their origin.  The magnetic phase multiplying a cooperon loop is proportional to the magnetic flux $\Phi$ through that loop, i.e. the product of the loop's area $A$ and the magnetic field strength $B$. Through the remainder of this text we will use atomic units where magnetic field has units of inverse area, the flux quantum $\Phi_0 = 2\pi \hbar/e = 2\pi$ is equal to $2\pi$, and  the magnetic flux is simply $\Phi = A B$.   In these units magnetic field and area are conjugate quantities, and a Fourier transform with respect to field $B$ results in a function of area $A$.

 Because  the area of cooperon loops can be much larger than the area of the unit cell or the area scale associated with scattering, the field-induced phase typically results in very sensitive dependence on magnetic field.  The strong sensitivity to small variations in field is a hallmark which allows experimentalists to easily distinguish the effect of the magnetic phase factor from other effects. Therefore experiments often smoothly and systematically vary magnetic field and  focus their attention on the sample's response to small changes, which is determined by the magnetic phase factor.

At a given Fermi level, in a particular sample, typically several cooperon loops contribute at the same time to the total conductance.  Each loop has its own chaos-determined trajectory with a corresponding area $A$, and therefore its contribution to the conductance oscillates with field strength at a characteristic periodicity $\Delta B = 2 \pi / A$ proportional to $1/A$.  The sum of many different signals from many Cooperon loops, each with their own period, produces a "random"-looking  dependence on $B$ which is however repeatable and provides a fingerprint which is unique to each sample and each Fermi level.  These quasi-random sample-specific signals are called universal conductance fluctuations (UCFs).

In this connection we recall that the quantum theory of conduction, as summarized in the Landauer-Buttiker formalism, states that there is a fundamental conductance quantum, $G_0 = e^2/ h$.  Conduction mediated by electron and hole charge carriers is naturally measured in these units, and under the right conditions is quantized. \cite{wharam1988one,beenakker1991quantum} This suggests that, at least under the right conditions, charge conduction is mediated by individual conducting channels, and that measurements of the conductance are equivalent to counting the integer number of channels through the sample that connect the two leads used to perform the measurement.

The standard perturbation theory of disordered conduction predicts that the size of Universal Conductance Fluctuations is generically of order $1 \, G_0$, even in samples whose conductance may be far larger than $G_0$.  [This is in the  absence of temperature effects, which tend to destroy quantum coherence, smooth the integer steps in the conductance, and multiply UCFs by a small exponential factor. ] This prediction, which actually happened only after experimental observation, was a significant surprise since  it had been expected that such samples would be far from the quantum limit and therefore quantum effects would be very small.  \cite{PhysRevB.30.4048,PhysRevLett.54.2692,PhysRevLett.54.2696,PhysRevLett.55.1622} 
As we have mentioned, the standard theory does not access UCFs directly, but instead calculates the second moment (standard deviation) of the conductance which is caused by UCFs. 
However, in combination with  predictions and observations of conductance quantization, these results suggest that UCFs are caused by specific diffuson or cooperon paths connecting the two leads in a sample, and that these paths are discrete and countable, resulting in an integer conductance when $B=0$, and in a superposition of cosines at non-zero $B$.

\subsection{Weak Localization and Antilocalization}
While the individual loops are effectively random, their statistical properties are not.  Cooperon loops with large areas occur less frequently than loops with small areas, because once a Cooperon has wandered far from its starting point it is less likely to return.   In a 2-D sample where motion occurs in a plane, a very precise probability distribution governs the statistics of Cooperon loop areas: the probability $P(A)$ of a loop with area $A$ is proportional to $1/A$.  \cite{sacksteder2018fermion,PhysRevB.61.13164}

In a trace of the conductance $G(B)$'s dependence on magnetic field, the $1/A$ statistics of Cooperon loop areas manifests as distinctive signature: the Hikami-Larkin-Nagaoka (HLN) logarithmic curve.  \cite{Hikami80} The quasi-random UCFs of the many individual Cooperon loops, superimposed on each other, add up to a logarithmic dependence on $B$.  The HLN logarithm varies slowly at large fields, but it accelerates at small fields, and would diverge logarithmically at zero field were it not for finite temperature and finite sample  size.   This extra logarithmic factor caused by quantum interference in Cooperons multiplies the classical conductance determined by diffusons.  In  materials where spin polarization relaxes slowly the logarithm   reduces the conductance and increases the resistance, and is called weak localization (WL).  In other materials such as topological insulators where spin polarization relaxes quickly the logarithm has the opposite effect, decreasing the resistance, and is called weak antilocalization (WAL).  The net effect is that a trace of the conductance's dependence on magnetic field shows quasirandom sample-specific universal conductance fluctuations (UCFs) at small changes in field, while at larger changes in field  these add up to the HLN logarithm which varies slowly at large fields but diverges very rapidly at small fields.

\subsection{Experimental Observation of Cooperon Loops}
In a recent paper we presented experimental measurements of conduction on  the surface of topological insulator samples, which manifested both the HLN logarithm and UCFs. \cite{kolzer2019phase}  By taking the Fourier transform of the magnetic field trace $G(B)$ we obtained its transform $G(A)$ which depends on area.  Aside from the average $1/A$ behavior, we found sharp peaks at specific values of the area $A$.  Each of these peaks  signals the existence on the sample's surface of a well defined  Cooperon loop with area $A$.   The Cooperons in these samples  had widths that were substantially smaller than the scale [width or height] of the Cooperon loops - otherwise the peaks in $G(A)$ would have smeared together and would have been impossible to resolve.  In other words, the observation of clear peaks at specific values of $A$ is clear evidence that in our TI samples the Cooperons did not extend throughout the TI surface like waves.  Instead the Cooperons followed well-defined paths, like particles.  
 
 These experimental results show that careful analysis of magnetic field dependence data from  an individual sample can reveal the details of the individual Cooperon loops in  that specific sample. We expect that this approach toward measurement and analysis of UCFs will be a powerful diagnostic tool for understanding conducting pathways in disordered materials.

\subsection{Magnetic Phase Effects in Wires}
The magnetic phase $\exp(\imath \Phi)$ can be examined in more detail in  wires constructed in such a way that conduction is permitted on the wire's surface but not in its interior.  This type of wire has been realized in heterostructures \cite{jung2008quantum,serra2009conductance,PhysRevB.89.045417}, graphene nanotubes \cite{stojetz2007competition}, and most recently in topological  insulator wires \cite{peng2010aharonov}.  In  these cases Cooperon loops on the wire surface are constrained to wind around the wire an integer number of times and may not penetrate into the bulk of the wire or complete a fractional number of windings. If  a magnetic field is oriented parallel to the axis of one of these wires, then the magnetic flux through the Cooperons loops in that wire is quantized.   The flux through each loop  is equal to the magnetic flux through the wire's cross-section, multiplied by the loop's winding number $n$, where $n$ is the number of times that the Cooperon loop revolves around the wire axis before coming back to its starting position on the wire.

This flux quantization has a very distinctive signature in the conductance $G(B)$: the resulting conductance signal is strictly periodic in $B$, with the period $\Delta B = 2 \pi/A$ inversely proportional to the wire cross-section $A$.  In other words, $G(B) = G(B+ 2 \pi /A)$  for all $B$.   Apart from the periodicity restriction, the conductance remains unconstrained by the wire geometry; the details of $G(B)$ are instead determined by the position of scattering centers, the same as UCFs in other materials.   The fundamental frequency $G(B) \propto \cos(1 \times 2 \pi A B)$, associated with an electron or hole making a single circuit around the wire, is called Aharonov-Bohm (AB) oscillations.  The first harmonic  $G(B) \propto \cos(2 \times 2 \pi A B)$, associated with a Cooperon  making a single circuit around the wire, is called Altshuler-Aronov-Spivak (AAS) oscillations.  In general $G(B)$ will be a superposition of AB oscillations, AAS oscillations, and every other winding number $G(B) \propto \cos(n \times 2 \pi A B)$.  An exponential dephasing factor cuts off winding numbers $n$ which exceed the ratio $l_\phi/P$ of the the dephasing length $l_\phi$ to the perimeter $P$. \cite{RevModPhys.59.755}

The results of the ensemble-averaged theory of disordered conduction are well known: while a clean [ballistic] sample will show AB oscillations i.e. the wire's fundamental frequency, these oscillations are mediated by single-electron [or hole] conduction, and are not robust against scattering.  Therefore in diffusive samples the fundamental frequency i.e. AB oscillations is suppressed. The  lowest visible frequency is therefore AAS oscillations, which are mediated by the Cooperon and therefore are robust against scattering. 

The above predictions are valid only for the ensemble average of the conductance of many samples.   We performed numerical simulations of conduction in single TI wires where the scattering length was comparable to or longer than the wire perimeter, but was much smaller than the wire length, putting each wire in the diffusive regime.   \cite{PhysRevB.94.205424}
We showed that  in a single wire [or nanotube, etc.] ALL winding numbers will be seen, including AB oscillations, AAS oscillations, and all higher winding numbers.    The weights of all winding numbers are quasirandom numbers, all of similar magnitude, with values determined by the positions of the scattering centers and by the Fermi energy.  In other words,  in single wires UCFs are dominant compared to any single-frequency oscillation such as AB and AAS oscillations. This prediction that  UCFs  are dominant, i.e. that all winding numbers will be seen with quasirandom weights of similar magnitude, is very generic for the simple reason that conduction on the disordered wire surface is thoroughly chaotic, giving every outcome and every winding number roughly equal probability.  In particular, the experimental observation of the AB fundamental frequency, vs. the AAS $n=2$ winding numbers, does not allow one to determine whether a particular wire is ballistic or diffusive, because UCFs cause both ballistic and diffusive wires to manifest both AB and AAS frequencies as well as all other harmonics.

Our finding of AB oscillations in the diffusive regime confirmed and extended a recent experiment   which found an AB signal in quite long wires  as long as the perimeter $P<l$ is less than than the scattering length $l$. \cite{dufouleur2015pseudo}  We found that if $P<l$  the AB signal depends periodically on $E_F$, and if $P>l$ its amplitude is random.   Our observation of AAS oscillations  in the ballistic regime verifies  theoretical work showing that  they can occur even in ballistic samples  as a consequence of constructive quantum interference between time-reversed circuits around the wire. \cite{kawabata1996altshuler,PhysRevB.57.6282,PhysRevB.69.205403,PhysRevB.70.161302}

It is only when the dephasing length is substantially smaller than the wire parameter that one is likely to observe only the fundamental AB oscillation frequency [already  damped by an exponential factor to have a magnitude well below $1 \,G_0$], because in this case the higher winding numbers are destroyed by dephasing.    \cite{RevModPhys.59.755}  Probably most articles on AB oscillations in TI nanowires belong in this regime \cite{hong2014one,jauregui2016magnetic, kim2016quantum}, although it is hard to  be certain because most do not report absolute scale of the conductance signal in units of $G_0$ and instead report it in percentages or without a scale, or alternatively report only the resistance not the conductance.   The most likely reasons for experiments operating in the regime where dephasing is strong are a tendency to self-select the experimental parameters including the temperature to get a clean cosine signal, and as well as the extra expense of reducing the temperature to the small values \cite{PhysRevLett.110.186806,cho2015aharonov} necessary to get dephasing lengths comparable to the wire perimeter.

\subsection{Experimental Direct Observation of Cooperon Winding Number}
In 2013 a Beijing experimental group reported an extraordinary result in an SnTe TI nanowire. \cite{safdar2013topological}  As the magnetic field was varied from 0 Tesla to about $2.3$ Tesla they observed  fourteen equally spaced oscillations in the conductance with a period of $\Delta B =0.165$ Tesla.  At 2.3 Tesla they observed an abrupt transition to oscillations with a period about twice as large, $\Delta B = 0.313$ Tesla.  Eleven of these oscillations were seen at field strengths reaching $6$ Tesla.  Both sets of oscillations were smooth i.e. similar to simple cosines without upper harmonics, and both sets had similar amplitudes. The oscillation amplitude decreased rapidly with increasing temperature, indicating that these oscillations were caused by quantum interference associated with the magnetic phase factor.

A similar result, but in a radial core/shell heterostructure wire, was reported by a Korean group in 2008. \cite{jung2008quantum} Like the Beijing group, the Korean group saw two different intervals of magnetic field, and in one interval the oscillations were twice as fast as in the other interval.  This data set's interpretation is a bit less clear cut, so for the remainder of this discussion we focus on the Bejing group's results.

There are several exceptional features of the Beijing data set. First, true AB oscillations would have maintained periodicity across the entire measurement range, as would have true AAS oscillations.  Instead the experimental signal shows a sharp  step-function-like transition at $B=2.3\,T$ which breaks the periodicity, signaling physics beyond the magnetic phase factor which is caused by loops around the surface of the wire.  The change in periodicity must have been caused by  some additional physical process which occurred at $B=2.3\,T$, such as a shift in Fermi level.  In disordered materials small shifts in the Fermi level are sufficient to completely change the paths of cooperons and diffusons, and therefore to change also the UCF fingerprint in the conductance.  Possibly the effect of magnetic field on the leads or on experimental equipment external to the sample  was not sufficiently controlled, causing a slight change in Fermi level and leading to the observed transition at $B=2.3\,T$.

Secondly:  As discussed earlier, when a single wire's perimeter is comparable to or smaller than  the scattering length,  the conductance will have components with all winding numbers, and each component will have  a quasirandom weight, resulting in a complex signal profile.  In contrast to this prediction, the experimental data shows instead a nice cosine profile at fields below $2.3\,T$, and another good cosine at higher fields.   In other words, a single well-defined winding number, probably $n=1$, is seen at $B = [2.3, 6] \,T$.  Another well-defined winding number, twice as large, probably $n=2$, is seen at $B = [0, 2.3] \,T$.

Thirdly, we consider whether temperature-induced dephasing could cause the observed simple cosine signals.   Dephasing    multiplies  each winding number by a small exponential factor, and the exponent in this factor is proportional to the winding number.  If the $n=1$ oscillations at $B = [2.3, 6] \,T$ are multiplied by a dephasing exponential, then the $n=2$ oscillations at $B = [0, 2.3] \,T$ must be multiplied by the square of that small number, and must be much smaller than the $n=1$ oscillations.  Instead the experimental data show that both sets of oscillations had a very similar amplitude.  This excludes temperature-induced dephasing  as the cause of the simple cosine profiles.

Fourthly, since temperature-induced dephasing cannot produce the observed well-defined winding numbers, we conclude that the cooperon loop responsible for the  $B = [2.3, 6] \,T$ signal really did follow a path around the TI wire which had a well-defined winding number $n=1$, i.e. it wound around the wire surface exactly once.    Similarly, the cooperon loop responsible for the $B = [0, 2.3] \,T$ signal wound around the wire surface exactly twice.

The presence of a single winding number signals that the cooperons on the surface of this TI wire followed particular well-resolved paths around the TI wire's surface.  The cooperons   did not extend  over the whole TI surface, since in this case there would have been no winding number.  The width of the cooperon trajectory, set by the scattering length, was substantially smaller than the perimeter of the wire.  In this sample the cooperon acted as a particle rather than a wave, following a well-defined loop around the wire.

\subsection{Experimental and Theoretical Recommendations}

Careful measurements of magnetic field dependence, coupled with Fourier transforms of this dependence, can give rich data on the areas of the loops contributing to conduction in the sample.  We recommend systematic emphasis on magnetoconductance measurements and their Fourier transforms, with finer resolution in field strength, more careful control of systematic and leads effects, variations of exploration of the angular orientation of the field, and careful experimental work on the form of the ultraviolet and infrared cutoffs on the loop area distribution.  This approach can give very detailed information about the geometry of the loops contributing to change transport. Consistent reporting of the magnitude of the conductance and its oscillations, as compared to the conductance quantum $G_0$, would also be very useful.  

It may be fruitful to look in diffusive samples  for other evidence of  pathways [followed by diffusons and cooperons] which have a finite width and depend sensitively on Fermi level, using perhaps a scanning tunneling microscope (STM)  or a superconducting quantum interference device (SQUID) microsocope.

For theorists there are a number of interesting questions, including understanding more precisely when cooperons and diffusons should act more like waves and when they should act more like particles.  One obvious need is to give experimentalists guidance on how many distinct Cooperon loops should be expected to be seen in a particular diffusive sample, and on what parameters control this.   Another need is theoretical calculations of the conductance in individual samples whose width and length  are both greater than the scattering length, combined with efforts to identify the paths taken by cooperons and diffusons.  In TI samples this may be difficult because of the cost of modeling the bulk, but perhaps for other materials the numerical cost will not be prohibitive.

 \begin{acknowledgments}  
 The Korean data set \cite{jung2008quantum} was first called to our attention around 2009 by Mahn-Soo Choi.  We thank him and Stefan Kettemann for the discussions at that time and for hospitality.   We also particularly thank Quansheng Wu, Yongqing Li, Jonas Koelzer, Thomas Schapers, and Alex Petrovic for collaboration and important discussions.
\end{acknowledgments}

\bibliography{Vincent}

\end{document}